\documentclass{sig-alternate-10pt}
\usepackage{subfig}

\usepackage{multirow}

\usepackage[usenames,dvipsnames]{color}
\usepackage{amsmath}
\usepackage{amsfonts}

\usepackage{times}
\usepackage{xspace} \usepackage{epsfig} 
\usepackage{amsmath}
\usepackage[hyphens]{url}
\usepackage{amsfonts} 

\usepackage{listings}
\usepackage{fancyvrb}
\VerbatimFootnotes

\newcommand{\comment}[1]{}
\newcounter{note}[section]

\usepackage{pifont}

\newcommand{\vyas}[1]{}
\newcommand{\seyed}[1]{}
\newcommand{\alig}[1]{}

\newcounter{packednmbr}

\newenvironment{packeditemize}{\begin{list}{$\bullet$}{\setlength{\itemsep}{0.5pt}\addtolength{\labelwidth}{-4pt}\setlength{\leftmargin}{\labelwidth}\setlength{\listparindent}{\parindent}\setlength{\parsep}{1pt}\setlength{\topsep}{0pt}}}{\end{list}}

\begin{document}

\title{Modeling Human Mobility and its Applications in Routing in Delay-Tolerant Networks: a Short Survey}

\author{Seyed Kaveh Fayazbakhsh\\Stony Brook University}

\maketitle 

\begin{abstract}

Human mobility patterns are complex and distinct from one person to another. Nevertheless,
motivated by tremendous potential benefits of modeling such patterns in enabling new 
mobile services and technologies, researchers have attempted to capture salient characteristics of 
human mobility.
In this short survey paper, we review some of the major techniques for modeling humans' co-location, 
as well as predicting human location and trajectory. Further, we review one of the most important application 
areas of such models, namely, routing in delay-tolerant networks.
\end{abstract}

\keywords{Human mobility, delay-tolerant networks, routing}

\section{Introduction}
The ability to model human mobility is key to developing various mobile technologies such as context-aware 
services, e-health, cloud computing, and delay-tolerant networking, to name a few. In this survey,
we review the state-of-the-art research on modeling humans' collocation and predicting people's locations and trajectories ($\S$\ref{sec:humanmobility}). 
Then we focus on the applications of these models in the specific area of routing in 
delay tolerant networks ($\S$\ref{sec:dtnrouting})---sparse wireless networks with limited connection opportunities. The choice
of delay tolerant networks as the application area is not random: mobile networks have been shown to be a vehicle for both collecting a wealth 
of information on human mobility and applying new services enabled by human mobility models. 

We deliberately avoid discussing technical details in order to provide an overview of the area. 
The interested reader is referred to the references for further details.
It is worth noting that a number of human mobility datasets are available in CRAWDAD~\cite{crawdad},
which is an online collection of many networking-related datasets and analysis tools.

\section{Human Mobility}
\label{sec:humanmobility}
In this section, we review the literature on human mobility models in three related categories:
inter-contact time (i.e., measuring how often two people are co-located) in $\S$\ref{subsec:inter}, location prediction 
(i.e., predicting where a person will be at time $t$ and for how long) in $\S$\ref{subsec:pred}, and
human mobility models (i.e., modeling the trajectory of a person) in $\S$\ref{subsec:traj}.

\subsection{Inter-Contact Time}
\label{subsec:inter}

\emph{Inter-contact time} is the duration between two consecutive times when
a given pair of mobile devices are co-located (i.e., they are near each other). Chaintreau et al.~\cite{ict1} observed that the distribution of the inter-contact times 
in a mobility dataset exhibits a heavy tail such as that of 
a power law distribution. This observation has played a key 
role in advancing the future research. Delving more into statistical properties 
of inter-contact times,  Karagiannis et al.~\cite{ict2} 
showed that before a certain time threshold, the CCDF 
(i.e., Complementary CDF given by $P (X > x)$) of inter-contact 
times follows the power law and after the threshold the CCDF resembles the 
exponential distribution. This observation was based on studying 
a number of GPS, GSM, WiFi, and Bluetooth datasets. Recently, these statistical 
characteristics of inter-contact times have been reconfirmed~\cite{newInter}.

\subsection{Location Prediction}
\label{subsec:pred}
Lee and Hou~\cite{lee} used a dataset of users' associations with access 
points (APs) and applied transient analysis of a semi-Markov model 
to design a timed and location prediction algorithm that can predict 
the future location of a user including both the future access points 
she is going to be associated with and her future associations' durations. 
Technically, a \emph{state} in this model represents an AP that the user 
is associated with at a given time instant. The input to the algorithm is 
the history of user-AP associations, and the output is $\phi_{ij}(k)$, which is
the probability of the event that the user will be in state $j$ after $k$ time intervals if she 
is currently in state $i$.

WhereNext~\cite{wherenext} is a classification-based scheme to learn the 
trajectory of a moving object based on the history of its movements. The 
performance of the classifier is evaluated over a dataset of 17,000 
cars equipped with GPS. In contrast to the previous work that can only predict
the next location of a user but not his/her arrival time and residency time 
(i.e., the interval of time spent at a location), NextPlace~\cite{nextplace} 
tries to estimate the duration 
of a visit to any given location and the time interval
between two subsequent visits to that location.

There are two classes of mobility models~\cite{chon}. The basic assumption of a 
\emph{location-dependent model} (e.g.,~\cite{lee}) is that people 
tend to remain at the same place for similar durations on each visit. 
A \emph{location-independent model} (e.g., ~\cite{nextplace}), on the 
other hand, uses temporal features without location information. 
Chon et al.~\cite{chon} analyzed these two classes of mobility models
using fine-grained mobility data. They deployed LifeMap, a mobility learning 
system to collect real user traces over a two-month 
period. LifeMap monitors the user's mobility every two minutes, 
using GSM, WiFi, and GPS. The system automatically recognizes visited 
places with room-level accuracy using WiFi fingerprinting. The basic 
idea is that radio signals from surrounding WiFi access points
(APs) are similar when a user is stationary at the same location. The authors have
drawn several conclusions on the advantages of
each of the two approaches to modeling mobility. Perhaps most notably, their
experiments showed that location-dependent models tend to have a higher
accuracy in predicting human's temporal behavior.

\subsection{Modeling Human Trajectory}
\label{subsec:traj}
Motivated by the fact that a realistic human trajectory (also referred to as human mobility) model
can be useful in networking-related simulation studies, Lee et al.
~\cite{slaw} developed a human mobility model. Past research had
shown that human mobility has certain statistical characteristics, namely:

\begin{packeditemize} 
 \item \textit{Truncated power-law flights and pause times}; where 
    a (human) flight is a straight line trip without any directional 
    change or pause.

 \item \textit{Heterogeneously bounded mobility areas}; people mostly
  move within their own confined areas of mobility, and
 different people may have widely different mobility areas.

 \item \textit{Truncated power-law inter-contact times} (see $\S$\ref{subsec:inter}).

 \item \textit{Fractal waypoints}; people are always more attracted to
  more popular places.

\end{packeditemize} 

Building on these characteristics, Slaw~\cite{slaw} presents a mobility model for 
mobile networks that can produce synthetic human mobility traces that possess all 
these statistical features.

Rhee et al.~\cite{levy}, in their seminal work, argued that despite certain well-understood characteristics of human 
mobility, no empirical evidence existed to prove the accuracy of the models founded
on them. They studied the mobility patterns of humans up 
to the scales of meters and seconds using mobility
track logs obtained from over 100 participants carrying GPS
receivers in five sites. They concluded that human walk
patterns involve statistically similar features to those observed in Levy
walks. These features include heavy-tail flights and pause-time
distributions and the superdiffusive nature of mobility.
Gonzalez et al.~\cite{nature} studied the trajectories of 100,000
mobile phone users whose positions were tracked for a 6-month
period. They found that the travel patterns of individual
users could be approximated by a Levy flight up to a certain
threshold distance. Moreover, they observed that
the individual trajectories are bounded beyond the threshold
distance; thus, large displacements, which are the source of the
distinct and anomalous nature of Levy flights, are statistically
absent. As a results, human trajectories were found to show a high
degree of temporal and spatial regularity.

A related question is to what extent human mobility is predictable? Song et al.~\cite{science}
studied a 3-month-long record capturing the mobility
patterns of 50,000 individuals. By measuring the entropy
of each individual's trajectory, they found a 93\% potential
predictability in user mobility.

Pu et al.~\cite{visual} took a different approach to analyzing
mobility of mobile phone users. They analyzed phone call
records of three million mobile phone users in a city over a
span of one year. Each phone call record usually contains the
caller and callee IDs, date and time, and the base station where
the phone calls are made. Using visualization techniques, they
classified users into distinct groups based on their mobility
patterns.

Tournoux et al.~\cite{accor} studied the motion of a population
of rollerbladers and observed a behavior called \emph{accordion
phenomenon}, which basically means people get close to and
far from each other with some harmonic delay. The authors
made use of this observation to tune the spray-and-wait~\cite{spray} routing protocol 
(see $\S$\ref{subsec:oppfor}).

\section{Routing in Delay-Tolerant Networks}
\label{sec:dtnrouting}

A Delay-Tolerant Network (DTN) is a sparse wireless network in which most of the time there does not exist a complete
path from a given source to its destination. Thus, conventional
routing schemes fail in DTNs. There is a considerable body
of literature devoted to developing efficient routing techniques
in DTNs, some of which utilize the properties of human mobility.

\subsection{Opportunistic Forwarding}
\label{subsec:oppfor}

Opportunistic forwarding methods are characterized by forwarding messages greedily or based on contact prediction as
nodes encounter each other. An initial attempt for routing in
DTNs was Epidemic Routing~\cite{epidemic} using which a node copies
the message to every other node it encounters that does not already have a
copy of the message. Message copies time out.
The hope is that at some point the destination node receives the message. 
Spray-and-wait~\cite{spray} departs from the epidemic method in that it
controls the number of copies of each message in the network.
In particular, a number $L$ of logical tickets are associated with
each message. Node $i$ copies a message to node $j$ that it
encounters only if the message owns $L > 1$ tickets or $j$ is the
destination. The new copy in $j$ will have $L_j = \lfloor \frac{L}{2} \rfloor $
tickets and $L_i = L -  L_j$ tickets will remain with the message in
$i$. MaxProp~\cite{maxprop} is based on prioritizing both the schedule
of packets transmitted to other peers and the schedule of
packets to be dropped. It uses several mechanisms to define the
priority based on which packets are transmitted and deleted.
MaxProp protocol uses a ranked list of the node's stored
packets based on a \emph{cost} assigned to each destination. The cost
is an estimate of delivery likelihood. In addition, MaxProp uses
acknowledgments sent to all nodes to notify them of packet
deliveries to prevent further transmission attempts. 

More recently proposed routing schemes account for node resource 
constraints. RAPID~\cite{rapid} makes the case that a contact may be too
short to transmit all packets, so it is important to determine
in what order packets should be forwarded. RAPID considers
the DTN routing problem as a resource management problem
in which various performance criteria such as average delay,
delivery deadline, and maximum delay can be incorporated.
Delegation Forwarding~\cite{delfor} exploits optimal stopping theory 
to decide whether an encountered node would be a good relay at
the moment of encounter. Similarly, optimal probabilistic forwarding (OPF)~\cite{opf} 
was designed under the assumption of long-term regularity of 
nodes mobility patterns as well as the assumption that each
node knows the mean inter-meeting times of all pairs of
nodes in the network. The forwarding rule (i.e., the decision on
whether to forward) depends on whether replacing the copy 
in node $i$ (the current node that contains the message)
with two new copies in node $i$ and node $j$ (a neighboring node) will
increase the overall delivery probability. Therefore, given a
certain set of constraints on the maximum number of forwarding attempts per
message, OPF maximizes the delivery probability of each message.

Using predict and relay (PER)~\cite{per},
nodes determine the probability distribution of future contact
times and choose a proper next hop in order to improve
the end-to-end delivery probabilities. The design is based on
two observations and explicitly takes advantage of human mobility models. 
First, nodes in a network within a social
environment usually move around a set of landmarks. Second,
in some social environments the node trajectory in time is
almost deterministic given the history of nodes mobility.

Spyropoulos et al.~\cite{spy} studied routing in DTNs whose
nodes are \emph{heterogenous}. Heterogeneity refers to the difference
in nodes classes (e.g. handheld devices, vehicles, and sensors).
The authors showed that heterogeneity would reduce the routing
performance when a typical routing scheme is used.
Consequently, they introduced the notion of \emph{utility} for each
network node that represents its capabilities that matter to the
routing strategy such as Most-Mobile-First (MMF), Spraying
or Most-Social-First (MSF) Spraying, etc. Crafting appropriate 
utility functions in designing utility-based protocols is effective in improving routing performance.

Encounter-based routing (EBR)~\cite{ebr} is based on the 
observation that the future rate of a node's encounters can be
roughly predicted by past data. This property is useful because
nodes that experience a large number of encounters are more
likely to successfully pass the message along to the final
destination than those nodes who only infrequently encounter
others.

Time-sensitive Opportunistic Utility-based Routing
(TOUR)~\cite{tour} considers \emph{time value} of messages: each message 
has a certain initial value that diminished with time. The goal is to 
maximize the remained time values of the messages when they get
to their destinations.

A common assumption in most DTN routing schemes is
that network nodes know their neighboring nodes. If there is
no infrastructure and devices must probe their environment to
discover other devices, an energy-aware contact probing
mechanism is required. If devices probe very infrequently, they might miss many of their contacts. On the other
hand, frequent contact probing might be energy inefficient.
Wang et al.~\cite{wang} study the problem of finding the optimal
probing time interval.

\subsection{Social-based Forwarding}
Research on social networks has been beneficial in designing DTN routing protocols. SimBet~\cite{simbet} is a routing
scheme that assesses similarity\footnote{Similarity of two nodes is defined as the number of neighbors these nodes have in common.}
to detect nodes that are
part of the same community, and betweenness centrality\footnote{Betweenness 
centrality of a node is defined as the fraction of shortest
paths between each possible pair of nodes going through this node.}
to identify bridging nodes that could carry a message from one
community to another. The decision to forward a message
depends on the similarity and centrality values of the newly
encountered node, relative to the current one: If the former
node has a higher similarity with the destination, the message
is forwarded to it; otherwise, the message stays with the most
central node. The goal is to fist make use of more central nodes
to carry the message between communities, and then use similarity 
to deliver it to the destination's community. BubbleRap~\cite{bub} 
is essentially similar to SimBet in that betweenness centrality
is used to find bridging nodes until the message reaches the
destination community. In contrast to SimBet, BubbleRap
explicitly identifies communities by a community detection
algorithm rather than doing so implicitly using similarity. Once arrived at
the destination community (i.e., the community in which the destination 
node is located), the message is only forwarded to other nodes
of that community: a local centrality metric is used to find 
increasingly better relay nodes within the community.

PeopleRank~\cite{people} is another social-based routing scheme
based on PageRank~\cite{page}. The intuition behind it is
that socially well-connected nodes tend to be more effective in 
message forwarding, simply because they meet more nodes. In PeopleRank, 
node $i$ forwards data to node $j$, which it has just met, if the rank
of $j$ is higher than the rank of $i$.
Nikolopoulos et al.~\cite{niko} challenged the suitability of using
centrality metrics in routing by conducting simulations. They
presented three reasons behind the observed inefficiency:
lack of destination awareness, use of estimated centrality
metrics due to practical limitations, and making over-simplifying assumptions in building
the contact graph.

Gao et al.~\cite{gao} presented the first  attempt toward 
understanding and formulating the problem of \emph{multicasting} in DTNs.
They utilized social networks concepts, including centrality
and social community, to improve the cost-effectiveness of
multicast in DTNs as compared to the case of using unicast 
schemes to carry out multicating. By modeling the problem 
using the unified knapsack problem, they developed forwarding
schemes that involve the forwarding probabilities to multiple
destinations simultaneously.

Gao and Cao~\cite{cao} considered forwarding data only to the
nodes that are interested in the data (called interesters). The
interesters of a data item are typically unknown in advance
at the data source because it is difficult for the data source
to acquire knowledge about the interests of other nodes in the
network \emph{a priori}. Therefore, the problem is different from multi-casting
and unicasting. They proposed a technique to relay selectively
based on node centrality values, and ensure that data items
are disseminated based on their popularities.

Hossmann et al.~\cite{hoss} argued that the prevalent pairwise
statistics (e.g., contact and inter-contact time distributions)
could not appropriately define a mobility model since a
mobility model should also reflect the macroscopic community 
structure of who meets whom. The authors observed that
communities are often connected by a few bridging links
between nodes who socialize outside of the context and
location of their home communities. They showed that it is
the social nature of bridges that makes them differ from
intra-community links. They argued that this observation
should be accounted for in the future schemes.

The social graph that represents frequency of contacts
between pairs of users and is used in social-based schemes
is typically an aggregation of contacts during a long time span.
Hossmann et al.~\cite{hoss2} argued that in social-based methods 
the mapping from the mobility
process generating contacts to the aggregated social graph is
more important than the routing algorithm in terms of their
respective effects on routing performance. The claim is verified
by their study of SimBet~\cite{simbet} and BubbleRap~\cite{bub}. Then, they
designed an online learning-based algorithm to infer the right level of aggregation
in building the contact graph. In a related work, Gao
and Cao~\cite{cao2} observed that the transient contact characteristics
of mobile nodes during short time periods in DTNs 
differ from their cumulative contact characteristics.

\subsection{Getting Selfish Nodes to Cooperate in Packet Forwarding}
To conclude the topic of DTN routing, we note that all the reviewed schemes 
(perhaps implicitly) make the assumption that network nodes are willing to cooperate in
forwarding the packets. However, without any incentives, selfish nodes would rather not 
collaborate and only take advantage of other cooperative nodes. 
Therefore, it appears an incentive-based mechanism should be
an integral part of any realistic DTN routing technique.\footnote{This is very similar to the 
problem of providing forwarding incentives in MANETs (see, for example,~\cite{sprite,me}).} We are
not going to focus on this issue and refer the interested reader
to~\cite{smart} as an early attempt to tackle this problem. Another related 
scheme has been proposed by Li et al.~\cite{li}. They
made the case that most people are socially selfish; i.e., they
are willing to forward packets for nodes with whom they have
social ties but not others, and such willingness varies with the
strength of social ties. They proposed the Social Selfishness
Aware Routing (SSAR) algorithm in which a forwarding node
is selected by considering both users' willingness to forward
and their contact opportunity, resulting in a better forwarding 
strategy than purely contact-based approaches. Despite these
few related works, it is somewhat surprising that providing 
cooperation incentives has been ignored in a majority of proposed
DTN routing protocols.
\section{Conclusions}

First, we reviewed the state-of-the-art methods of modeling human mobility.
Even though human mobility is inherently
complicated to capture, even relatively simple models that incorporate a few
of its key characteristics have been useful in 
advancing research on mobile networking, among other fields. Then, we
summarized the existing methods of routing in DTNs, which  
significantly benefit from human mobility models.

{
\bibliographystyle{abbrv}
\bibliography{./bibs/seyed,./bibs/amin}
}

\end{document}